\documentstyle[aps,preprint,graphicx]{revtex}
\tighten

\newcommand{\TCE}{{\tilde{\cal E}}}
\newcommand{\TE}{\tilde E}
\newcommand{\TG}{\tilde \Gamma}

\begin{document}

\title{Fano resonances in the overlapping regime}

\author{A. I. Magunov{$^{1,2}$}, I. Rotter{$^{1}$}, and
S. I. Strakhova{$^{3}$}}

\address{$^1$
Max Planck Institute for the Physics of Complex
Systems,  D-01187 Dresden, Germany \\
$^2$
General Physics Institute, Russian Academy of Sciences,
119991 Moscow, Russia \\
$^3$ Skobeltsyn Institute of Nuclear Physics, Moscow State
University, 119992 Moscow, Russia}

\date{\today}

\maketitle

\begin{abstract}
The  line shape of resonances in the overlapping regime is studied
by using the eigenvalues and eigenfunctions of the effective
Hamiltonian of an open quantum system. A generalized expression
$\tilde q_k(E)$ for the Fano parameter of the resonance state $k$
is derived that contains the interaction of the  state $k$ with
neighboured  states $l\ne k$ via the continuum. It is energy
dependent since  the coupling coefficients between the  state $k$
and the continuum show a resonance-like behaviour at the energies
of the neighboured  states $l\ne k$. Under certain conditions, the
energy dependent $\tilde q_k(E)$ are equivalent to the generalized
complex energy independent Fano parameters that are introduced by
Kobayashi et al. in analyzing experimental data.  Long-lived
states appear mostly  isolated from one another  in the cross
section, also when they are overlapped by short-lived resonance
states. The $\tilde q_k(E)$ of  narrow resonances  allow therefore
to study the complicated interplay between different time scales
in the regime of overlapping resonance states  by controlling them
as a function of an external parameter.

\end{abstract}


The line shape of resonances is studied in many different physical systems.
A resonance has a symmetrical Breit-Wigner line shape  when  it
is far in energy  from other resonances and from particle decay
thresholds, and when there is no
interference with any smooth background. Such a situation is seldom met
in realistic systems. The interference with an energy independent background
is taken into account in the approach  suggested by Fano
\cite{Fano} for the description of autoionizing atomic states.
The parametrization
of the cross section can explain the asymmetry  of the line shape of
isolated resonances. This method was extended
later to the  many-channel problem and to the overlapping between broad and
narrow resonances, see \cite{miess,friedrich} and the
recent papers \cite{tabanli}.

Deviations from the Breit-Wigner resonance form  appear also when the width of
the resonance state itself is energy dependent. This happens, above all,
in the neighbourhood of particle decay thresholds.
Under certain conditions, the Breit-Wigner
resonance shape turns over even into a cusp \cite{ro91}.
Furthermore, the tail of bound states (with position below the
first decay threshold) can be seen in the cross section at positive
energy and may even interfere with resonance states lying at these
positive energies \cite{rep}.

Energy dependent effects may appear, however, also far from thresholds when
the individual resonance states start to overlap \cite{rep}. Here,
interferences between the different resonance states having different
lifetimes, as well as with a smooth background may cause altogether a
complicated behaviour of the cross section in the neighbourhood of
the narrow resonance states. Experimental studies are performed on
microwave cavities as a function of the degree of opening of the cavity
\cite{stm}. In atoms, laser-induced structures
are studied theoretically in the non-Hermitian Hamiltonian approach
\cite{marost}. In these cases, the line shape can be controlled by means of
the parameters of the laser field. A systematic experimental study
is not performed, up to now.

The situation is especially complicated when the
narrow resonance states are well separated from one another in the cross
section and, nevertheless, overlapped by one or more broad ones
and, furthermore, the cross section contains a smooth (energy-independent)
part. The line shape of these resonances is   studied in detail
experimentally and theoretically in atoms with Rydberg series overlapped
by an intruder, e.g. \cite{conlane}  and the textbook \cite{friedrich}.

Another example are the
neutron resonances in heavy nuclei the line shape of which is not studied,
up to now. They represent a realistic
example of a Gaussian orthogonal ensemble. These resonances do not decay
according to an exponential law as shown theoretically
\cite{hardit} as well as experimentally \cite{harditexp}. Thus, their line
shape cannot be of Breit-Wigner or Fano type with energy-independent
coupling coefficients between resonance states and continuum
although the resonances are well separated in energy from one another.

Recently, the line shape of resonances as well as  phase jumps of the
transmission amplitudes have been discussed  in experiments on
electron transport through  mesoscopic systems
\cite{heiblum,madhavan,gores,kob}. The advantage of these
experiments consists in the fact that   the key parameters are
tuned, and the interference leading to Fano resonances is studied
in greater detail. In \cite{kob}, the conductance through a
quantum dot in an Aharonov-Bohm interferometer is controlled by
varying the strength of the magnetic field. The authors claim that
the Fano parameter $q$ has to be extended to a complex number. The
physical meaning of this result remains an open problem.

In order to understand the phase jumps, the effects of the signs
of the dot-lead matrix elements onto the appearance of
transmission zeros and the phases of the transmission amplitudes
have been studied theoretically, e.g. \cite{silva,imry,kim}.
These studies do not use the eigenfunctions of the Hamiltonian of
the system in calculating the coupling matrix elements. Such a
study is justified by the fact that the profile of the Fano
resonances is independent of the special properties of the system,
indeed \cite{Fano}. However, the simple relation between the
standard Fano parameters and the coupling coefficients between
system and environment holds only in the regime of non-overlapping
resonances. In the overlapping regime, the coupling coefficients
are energy dependent \cite{rep,ro03}.

The line shape of
resonances in the overlapping regime is  studied in \cite{marost} for
laser-induced structures in atoms. The formalism used is the effective
Hamiltonian describing  open quantum systems.
It can be applied to the study of the resonance phenomena in the
non-overlapping regime as well as in the overlapping regime  \cite{rep}.
The formalism of the effective Hamiltonian can be used also
for the description  of quantum dots.
It gives reliable results not only when the
dot-lead matrix elements are small but also when they are large
\cite{rep,stm,saro}. An interesting feature, proven experimentally \cite{stm},
is that narrow resonances appear for both small and large  dot-lead (or
cavity-lead) coupling coefficients. In the last case, they coexist with broad
resonances and appear mostly as dips in the transmission \cite{saro,napirose}.

It is the aim of the present paper to study the line shape of
resonances in the overlapping regime on the basis of the formalism
of the effective Hamiltonian ${\cal H}$. This Hamiltonian  appears
in the subspace of discrete states after embedding it into the
continuum (subspace of scattering states). It contains the
Hamiltonian of the corresponding closed system as well as the
coupling matrix elements between system and environment. It is
non-Hermitian and its eigenvalues $\tilde{\cal E}_k$ and
eigenfunctions $\tilde\Phi_k$ are complex and energy dependent.
The eigenvalues provide the poles of the $S$ matrix and the
eigenfunctions are used for the calculation of the coupling matrix
elements between system and environment, i.e. for the numerators of
the $S$ matrix in pole representation. The formalism  is described
in detail in the recent review \cite{rep}.

In the overlapping regime, different time scales exist
simultaneously due to the different crossings of the resonance
states in the complex plane that are avoided, as a rule
\cite{rep}. The avoided  crossings can be traced, as a function of
a certain parameter, in  the trajectories of the eigenvalues
$\TCE_k = \TE_k - \frac{i}{2} \TG_k$ of the non-Hermitian
effective Hamiltonian ${\cal H}$. Characteristic of the motion of
the poles of the $S$ matrix are the following generic results
obtained for very different systems in the overlapping regime: the
trajectories of the $S$ matrix poles avoid crossing with the only
exception of exact crossing when the $S$ matrix has a double (or
multiple) pole. At the avoided crossing, either level repulsion or level
attraction occurs. The first case is caused by a predominantly
real interaction between the crossing states and is accompanied by
the tendency to form a uniform  time scale of the system. Level
attraction occurs, however, when the interaction is dominated by
its imaginary part arising from the coupling via the continuum. It
is accompanied by the formation of different time scales in the
system: while some of the states decouple (partly) from the
continuum  and become long-lived (trapped), a few of the states
become short-lived and wrap the long-lived ones in the cross
section. The dynamics of quantum systems at high level density is
determined by the interplay of these two opposite tendencies
\cite{rep}. The Fano parameters reflect this interplay as will be
shown in the following.

Let us recall the  $S$ matrix
for an isolated resonance state $k=1$ in the one-channel case,
\begin{eqnarray}
S(E) = 1 -  \, i \; \frac{\tilde W_1}{E- \tilde{\cal E}_1} \; .
\label{brwi1}
\end{eqnarray}
Here, $\tilde{\cal E}_1 = \tilde E_1 - \frac{i}{2} \tilde \Gamma_1
$ is the complex energy of the resonance state (eigenvalue of the
effective Hamiltonian)  and $\tilde E_1 $ and $\tilde\Gamma_1$ are
its position in energy and width, respectively. The  $\tilde W_1$
are related to the coupling matrix elements between the resonance
state and the continuum  that are calculated by means of the
eigenfunctions of the effective Hamiltonian.  From the unitarity
of the $S$ matrix follows $\tilde W_1 = \tilde \Gamma_1$ for an
isolated resonance state, since Eq. (\ref{brwi1}) can, in the
one-resonance-one-channel case, then be written as
\begin{eqnarray}
S(E)&=&
 \frac{E- \TCE_1^*}{E- \TCE_1} \; .
\label{brwi2}
\end{eqnarray}
The $S$ matrix (\ref{brwi2}) is unitar. In the following, we assume that
$\TCE_1$ is nearly constant inside the energy range of interest.

The unitar representation of the $S$ matrix in the one-channel case with
two  resonance states ($k = 1, \, 2$)  and a smooth
reaction part can be written as
\begin{eqnarray}
S(E)&=& \exp (2i\delta ) \cdot \; \frac{E-\TCE_1^*}{E-\TCE_1}
\cdot \frac{E-\TCE_2^*}{E-\TCE_2} \label{li0}
\\[.3cm]
&=& \exp \{ 2 i \big(\delta + \delta_1(E) + \delta_2(E)\big) \} \;
. \label{li1}
\end{eqnarray}
Here,  $\delta$ is the phase shift related to the smooth part, and
\begin{eqnarray}
\delta_k(E)
& = & - \; {\rm arccot} \; \varepsilon_k
\label{li2}
\end{eqnarray}
with
\begin{eqnarray}
\varepsilon_k = \frac{2\, (E-\TE_k)}{\TG_k} \, .
\label{li5}
\end{eqnarray}
Using these expressions, the cross section reads
\begin{eqnarray}
\sigma(E) & = & 2 \, \big( 1- {\rm Re}\, S(E)\big)
 = 2\big[ 1-\cos(2\delta+2\delta_1(E)+2\delta_2(E)) \big] \nonumber \\
& = & 4\, \sin^2 \, \big( \delta + \delta_1(E) + \delta_2(E) \big)
\, . \label{li3}
\end{eqnarray}
In the vicinity of the energy $\TE_1$ of the  resonance state 1
one gets from (\ref{li3})
\begin{eqnarray}
\sigma(E)
 & = & 4\; \big[ \sin
 \eta(E)\cos\delta_1(E)+\sin\delta_1(E)\cos\eta(E)\big]^2
 \nonumber \\
& = & 4\; \sin^2 \eta(E)
\frac{[\cot\eta(E)+\cot\delta_1(E)]^2}{\sec^2\delta_1(E)}
\nonumber \\
 & = & 4\; \sin^2 \eta(E) \; \frac{\big(\tilde q_1(E) +
\varepsilon_1\big)^2} {\varepsilon_1^2 + 1}
\label{li4}
\end{eqnarray}
where $\eta(E) \equiv \delta + \delta_2(E)$ and
\begin{eqnarray}
\tilde q_1(E) = -\cot \eta(E) = \frac{\varepsilon_2 \cot\delta +
1}{\cot\, \delta - \varepsilon_2  }  \, .
\label{li6}
\end{eqnarray}

Eq. (\ref{li4}) is the Fano representation of the cross section in
the neighbourhood of the energy $\TE_1$ of the  resonance state 1
with interference contributions from the resonance state 2 and the
smooth reaction part. The Fano parameter $\tilde q_1$ is an energy
dependent function. It is $\tilde q_1(E) = \varepsilon_2$ when
$\delta \to 0$, and the energy dependence plays a role only for
overlapping resonance states. For non-overlapping resonance states
1 and 2  (i.e., $\varepsilon_2\to \pm\infty$ at the energy $E$ of
the resonance state 1) follows $\tilde q_1=-\cot\delta$
independently of $E$ in agreement with \cite{Fano}. In the
other extreme case of a very broad resonance state 2 ($\varepsilon_2\to
0$) $\tilde q_1=\tan\delta$ is also independent of $E$ but differs
from the foregoing case by the  additional phase $\pi /2$. It is
easy to generalize (\ref{li4}) to include the contributions from
additional resonance states
($\eta(E)=\delta+\sum_{k>1}\delta_k(E)$).

The energy dependence of the $\tilde q$ is caused by the energy dependence
of the coupling coefficients between system and continuum as will be shown
in the following.
Assuming $\delta \to 0$, two resonance states and one common channel,
the pole representation
\begin{eqnarray}
S(E) = 1 \, - \, i \, \sum_{k=1,2} \frac{U_k}{E-\TCE_k}\
\label{more4}
\end{eqnarray}
of the $S$ matrix can be derived from (\ref{li1}) in different ways.

\vspace{.2cm}

(i) $U_k = W_k$ are energy independent,
\begin{eqnarray}
 W_k & = & \TG_k  \Big( 1 - \,i
\frac{\TG_{l\ne k}}{\TCE_{l\ne k} - \TCE_k}
 \Big) \, .
\label{more5}
\end{eqnarray}
In deriving (\ref{more5}) from (\ref{li0}),
the standard representation of the term
$$\frac{\tilde \Gamma_1 \tilde\Gamma_2}{
(E-\tilde {\cal E}_1)(E-\tilde {\cal E}_2)}
=\frac{\tilde \Gamma_1 \tilde \Gamma_2}{
(E-\tilde {\cal E}_1) (\tilde{\cal E}_2 - \tilde{\cal E}_1)} -
\frac{\tilde \Gamma_1 \tilde \Gamma_2}{
(E-\tilde {\cal E}_2)(\tilde{\cal E}_2 - \tilde{\cal E}_1)}$$ is
used. The cross section reads
\begin{eqnarray}
\sigma(E) = \sigma_{a1}\,
\frac{(q+\varepsilon_1)^2+A_1}{\varepsilon_1^2+1}
+\sigma_{a2}\,
\frac{(q+\varepsilon_2)^2+A_2}{\varepsilon_2^2+1}
+\sigma_b
\quad
\label{morecs}
\end{eqnarray}
with the parameters
\begin{eqnarray}
q & = & \varepsilon_{12} = \frac{2(\TE_1-\TE_2)}{\TG_1-\TG_2}
\qquad A_k=\frac{\TG_k}{\TG_{l\neq k}}(\varepsilon_{12}^2+1)+
2(1-\varepsilon_{12}^2)
\label{more9}
\end{eqnarray}
and
\begin{eqnarray}
\sigma_{ak} =
\frac{4\TG_{l\neq k}}{(\TG_k-\TG_l)\; (\varepsilon_{12}^2+1)}
\qquad
\sigma_b = \frac{4}{\varepsilon_{12}^2+1}
\label{more9a}
\end{eqnarray}
that are all energy independent. The parametric expression
(\ref{morecs}) for the cross section is similar to that given in
\cite{FanoCooper} for the case of one resonance state coupled to
several decaying channels. The parameter $q$ can be considered as
Fano parameter. In our case of one resonance state coupled to
another one, the contributions  to each resonance state in
(\ref{morecs}) are modified by the additional  term  $A_k$.
According to  (\ref{more9a}), it is
$\sigma_{a1}+\sigma_{a2}+\sigma_b=0$, and at least one of the
parameters $\sigma_{ak}$ is negative. This parameter has no
physical meaning at all. The two parameters $q$ and $A_k$ are not defined at
a double pole of the $S$ matrix since they are singular when the double pole 
is approached  along the real energy axis. 
That means also the $W_k$ have, in this case, no physical meaning.
The singularities in (\ref{more9}) and
(\ref{more9a}) appearing in approaching  a double pole of the $S$
matrix  cancel each other in the expression for the cross section.

For overlapping resonance states with $\TG_2\gg\TG_1$ and
$|\TE_1-\TE_2|<\TG_2/2$, it holds $\varepsilon_{12}^2<1$. In this
case, the $A_k$ are positive and it is possible to introduce
complex energy independent Fano parameters
\begin{eqnarray}
q_k\equiv q+iA_k^{1/2} 
\label{more10}
\end{eqnarray}
in the equation (\ref{morecs})  for  the cross section,
\begin{eqnarray}
\sigma(E)=\sum_{k=1,2}\sigma_{ak} \frac{|q_k+\varepsilon_{k}|^2}{
\varepsilon_{k}^2+1}+\sigma_b
\label{more10a} \; .
\end{eqnarray}
This is in analogy to the result obtained in \cite{kob} from the analysis
of experimental data. It is also similar to an expression for the Fano
parameter obtained in a completely different approach \cite{koenig}.

\vspace{.2cm}

\begin{figure}
\begin{center}
\begin{minipage}{0.7\textwidth}
\includegraphics[width=\textwidth]{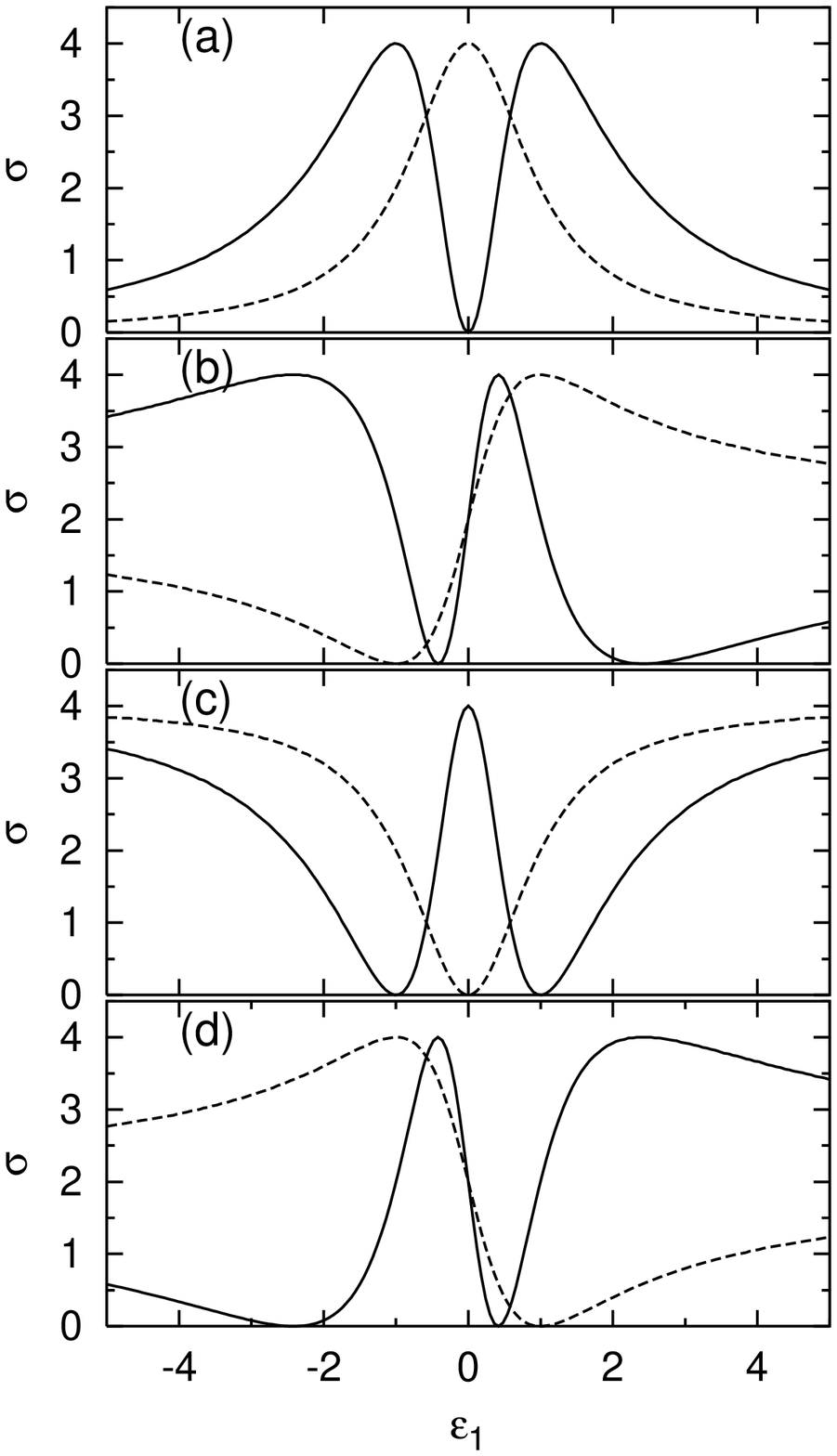}
\end{minipage}
\end{center}
\caption{
The  cross section in the neighbourhood of a
double pole of the $S$ matrix ($\TE_1=\TE_2$ and $\TG_1 = \TG_2$).
The direct scattering phase
$\delta$ is: (a) $\delta=0$,  (b) $\delta=\frac{1}{4}\pi$,
(c) $\delta=\frac{1}{2}\pi$, (d) $\delta=\frac{3}{4}\pi$.
The dashed curves correspond to the case of the two resonance states
without any interaction between them.
}
\label{fig1}
\end{figure}

(ii)  $U_k = \tilde W_k$ are energy dependent,
\begin{eqnarray}
\tilde W_k & = & \TG_k \; \Big( 1 - \,i \,
\frac{\TG_{l\ne k}}{2 E-  \tilde{\cal E}_k - \tilde{\cal E}_{l\ne k}}
\Big) \, . \quad
\label{more15}
\end{eqnarray}
In deriving (\ref{more15}), the non-standard representation of the term
$$\frac{\tilde \Gamma_1 \tilde\Gamma_2}{
(E-\tilde {\cal E}_1)(E-\tilde {\cal E}_2)}
=\frac{\tilde \Gamma_1 \tilde\Gamma_2}{
(E-\tilde {\cal E}_1) (2E-\tilde{\cal E}_1 - \tilde{\cal
E}_2)} +
\frac{\tilde \Gamma_1 \tilde\Gamma_2}{
(E-\tilde {\cal E}_2) (2E-\tilde{\cal E}_1 -
\tilde{\cal E}_2)} $$ is used.  The $\tilde W_k$ have a physical
meaning also in the overlapping regime. They are the coupling
coefficients between the states of the system and the continuum
that are calculated for realistic systems in the framework of a
unified description of structure and reaction aspects \cite{rep},
i.e. by means of the eigenfunctions of the effective Hamiltonian.
As can be seen from (\ref{more15}), the energy dependence of
$\tilde W_k$ appears due to its resonance behaviour at the energy
of another resonance state $l\ne k$. This is in complete agreement
with (\ref{li6}) for the energy dependent Fano parameter $\tilde
q_k(E)$: near the energy of the resonance state 1, the cross
section is given by (\ref{li4}) with the Fano parameter
(\ref{li6}).  Furthermore, the $S$ matrix goes over smoothly  in
\begin{eqnarray}
S & = &
1 - 2\, i \, \frac{\TG_d}{E-\TE_d + \, \frac{i}{2} \, \TG_d} -
\frac{\TG_d^2}{(E-\TE_d + \, \frac{i}{2} \, \TG_d)^2} \qquad
\label{brwi6}
\end{eqnarray}
when $\TE_1 \to \TE_2 \equiv \TE_d$ and $\TG_1 \to \TG_2 \equiv \TG_d$
(double pole of the $S$ matrix, see \cite{newton}).
The second term corresponds to the usual linear term while the third term is
quadratic.  The interference between these two parts is
illustrated in Fig. \ref{fig1} where the cross section is shown for the case
of two resonance states with equal positions $\TE_d$ and widths $\TG_d$,
coupled to one channel,  for different $\delta$.
The asymmetry of the line shape of both peaks is,
in the case $\delta=0$ (Fig. \ref{fig1}.a), caused solely by the
overlapping of the two resonance states. For comparison, the cross section
with the two resonance states without any coupling between 
them is also shown in Fig. \ref{fig1} (dashed curves).

\vspace{.2cm}

The energy independent coupling coefficients $W_k$, Eq. (\ref{more5}),
lose their physical meaning in the overlapping regime, since here
the energy dependence of
the coupling coefficients of the resonance states to
the continuum  can not be neglected (see
\cite{ro03} and \cite{rep} for numerical
examples).  It is different from that of the $\TG_k$ even in the one-channel
case  due to the biorthogonality of the
wave functions of the resonance states \cite{rep}. Nevertheless,
the two representations of the
$S$ matrix (\ref{more4})  with energy
independent $W_k$ and energy dependent $\tilde W_k$ are equivalent in almost
all cases. At the double pole of the $S$ matrix, however, the
$W_k$ have a singularity while the $\tilde W_k$ behave smoothly.
Since $\sigma (E) = |1- S(E)|^2$, the energy independent Fano parameters
can be expressed by $W_k$ and the energy dependent ones  by $\tilde W_k$.
That means, also the energy independent Fano parameters lose their physical
meaning in the overlapping regime.

As can be seen from (\ref{li6}), the energy dependent Fano parameter
$\tilde q_k(E)$
contains the influence from other resonance states and the background
without any contribution from its own $\TG_k$. Also
$\TE_k$ is not  directly involved in the
expression for the $\tilde q_k(E)$.
The $\tilde q_k(E)$ allows therefore to analyze, in a very transparent manner,
the mutial influence of resonance states in the
overlapping region in the neighbourhood of the resonance state $k$.
This is in contrast to the
energy independent Fano parameters $q$ that are more complicated
for an interpretation of the data. According to (\ref{more9}), they contain
the contributions from neighbouring resonance states
together with their own parameters.

In the multi-channel case, the matrix elements $S_{cc'}$ are of the same
structure as those  given in (\ref{more4}) for the one-channel case.
The coupling coefficients
\begin{eqnarray}
U_k^{cc'}=
\tilde W_k^{cc'} \equiv (\tilde W_k^c)^{1/2}\; (\tilde W_k^{c'})^{1/2}
\label{more25}
\end{eqnarray}
in  $S_{cc'}$ are, generally, complex and energy dependent \cite{rep}.
In the two-resonance case,  the $S$ matrix elements (with $\delta = 0$)
are
\begin{eqnarray}
S_{cc'}&= & \delta_{cc'} - i
\sum_{k=1,2} \frac{(\tilde W_k^c)^{1/2} \; (\tilde W_k^{c'})^{1/2}}{E-\TCE_k}
- \, i \, X_{12}^{cc'} \, . \qquad
\label{more26}
\end{eqnarray}
Here, the $X_{12}^{cc'}  \propto  1/\big((E-\TCE_1)(E-\TCE_2)\big) $
can be parametrized by means of energy dependent as well as
by energy independent
parameters in the same manner as in the one-channel case.
The energy independent parameters
have, however, no physical meaning at all in the overlapping regime.

\begin{figure}
\begin{center}
\begin{minipage}{0.65\textwidth}
\includegraphics[width=\textwidth]{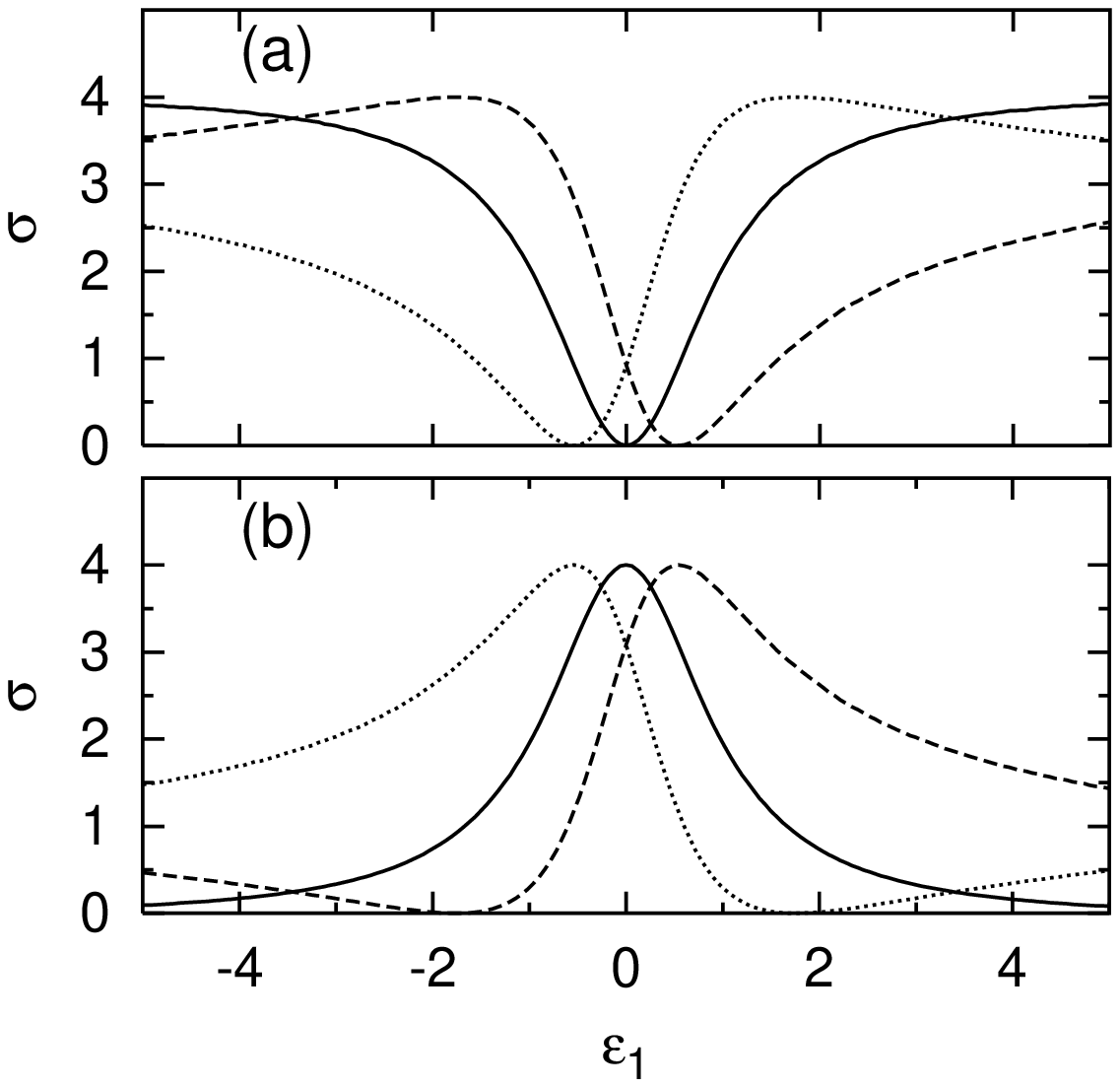}
\end{minipage}
\begin{minipage}[tr]{0.61\textwidth}
\includegraphics[width=\textwidth]{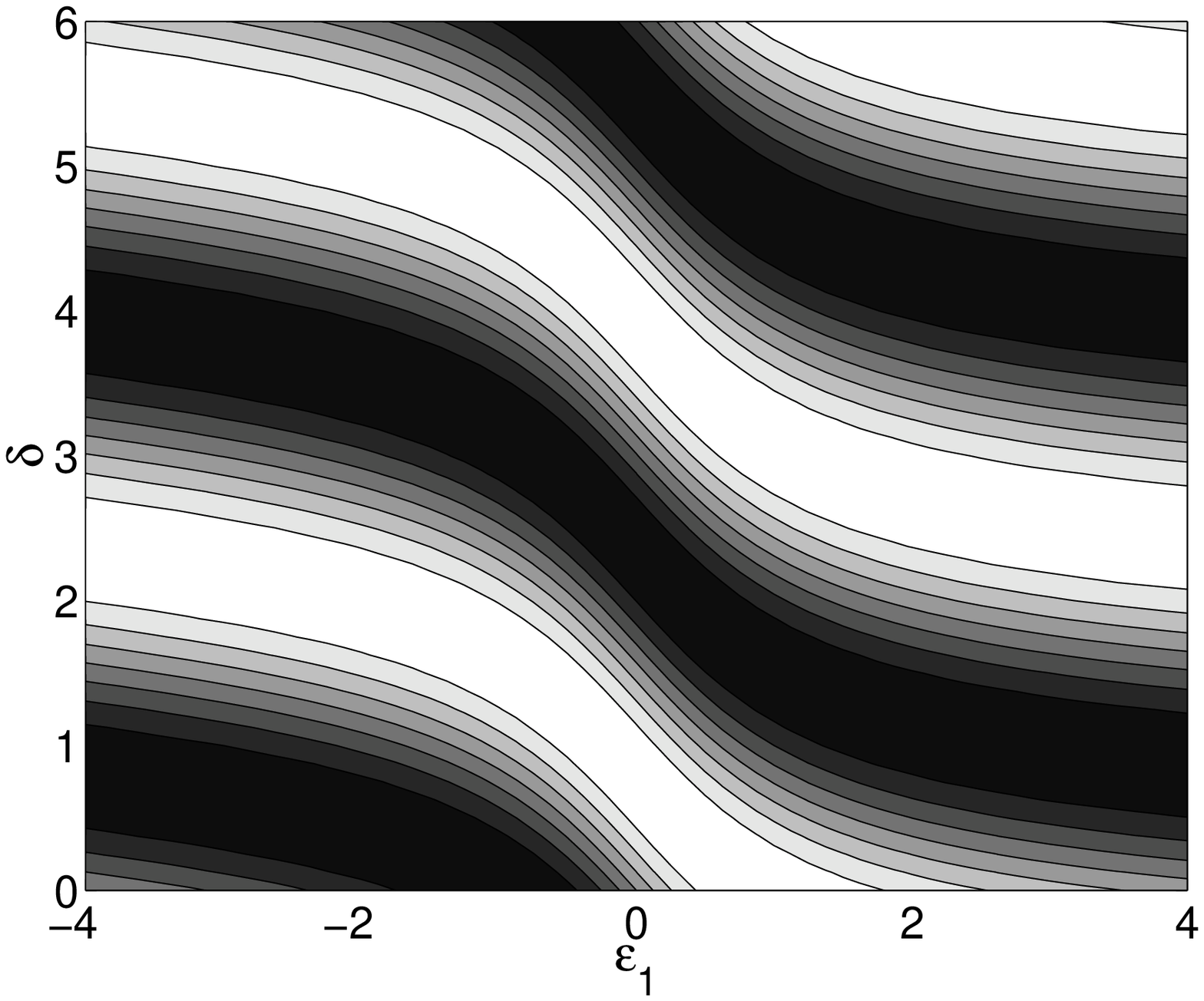}
\end{minipage}
\end{center}
\caption{ The cross section in the vicinity of the resonance 1 for
 $\TG_1 = 0.1 \, \TG_2, \; \TE_2 - \TE_1 = 0.5\,\TG_2 $.
Top: (a)  $\delta = \delta_0= -{\rm
arctan}\,\varepsilon_2(E=\tilde E_1)$ (full curve), $\delta =
\delta_0-0.5$ (dashed curve),  $\delta = \delta_0+0.5$ (dotted
curve). (b) The same as in (a), but $\delta_0={\rm
arccot}\,\varepsilon_2(E=\TE_1)$. Bottom: The contour plot of the
cross section. } \label{fig2}
\end{figure}

We discuss now the behaviour of the $\tilde q_k(E)$ in more
detail. It follows immediately from  (\ref{li6}) that $\tilde
q_1(E)=0$ at the energy
\begin{eqnarray}
E=\TE_2-\frac{1}{2} \, \TG_2 \, \tan\,\delta \; . \label{new17}
\end{eqnarray}
When this condition is fulfilled in the neighbourhood of $\TE_1$
and $\TG_1 \ll \TG_2$, then the narrow resonance state appears as
a window-type resonance (dip) in the cross section ($q$ reversal
\cite{lane}), see Fig. \ref{fig2}.a, full curve. It occurs,
however, as a Breit-Wigner type resonance when
\begin{eqnarray}
E= \TE_2 + \frac{1}{2} \, \TG_2 \, \cot\,\delta \label{new18}
\end{eqnarray}
is fulfilled near  $\TE_1$, see Fig. \ref{fig2}.b, full curve. In
Fig. \ref{fig2},  the narrow resonance 1 is  overlapped by the
broad resonance 2 ($\TG_1 = 0.1 \, \TG_2, \; \TE_2 - \TE_1 =
0.5\,\TG_2 , \; \TG_1 + \TG_2 = 1.1 \, \TG_2$) and interferes with
the smooth reaction part. Fig. \ref{fig2} remains almost unchanged
by varying $\TE_2$ and $\TG_2$ and correspondingly $\delta_0 $, as
long as $\TG_2 \gg \TG_1$. We underline that the narrow resonance
1 appears in the cross section as an isolated peak in spite of the
overlapping of the resonance state 1 with the broad resonance
state 2.  The reason is that Eq. (\ref{new18}) is fulfilled near
the energy of the narrow resonance, but not Eq. (\ref{new17}), due
to the interference with the smooth part.

The contour plot of the cross section (Fig. \ref{fig2} bottom)
shows that its resonance structure  varies periodically with the
phase $\delta$ of the background for fixed energies $E$.
Accordingly, the asymmetry of the line shape varies between
$\tilde q_1 \to - \infty$ and $\tilde q_1 \to + \infty$
periodically (as can be seen also  directly from the analytical
expression (\ref{li6}) for $\tilde q_1$). Such a behaviour is
observed experimentally in the conductance peaks through a quantum
dot as a function of the strength of the magnetic field \cite{kob}
(that has obviously a small influence  on the position and width
of the narrow resonances).
The complex Fano parameters introduced in \cite{kob} from a fit of
the data, simulate the overlapping of the studied narrow resonance
state by a broad resonance state and the phase dependence of
$\tilde q_k(E)$ in the overlapping regime. This conclusion follows
from the numerical study shown in Fig. \ref{fig2} as well as from
the analytical study resulting in Eqs. (\ref{more10}) and
(\ref{more10a}) for this case. 

In both representations, Eqs. (\ref{li6}) and  (\ref{more10}), 
the  Fano parameter is generalized when applied to resonances in
the overlapping regime.  Both representations are equivalent
in most cases. The difference between both 
generalizations can be seen best when applied  
to an analysis of the data in the vicinity of the energy of one of the 
resonance states. The energy independent complex
parameter (\ref{more10}) contains the energies and widths of both 
resonance states, so that it is difficult to receive spectroscopic
information. The energy dependent value   (\ref{li6}), however, 
contains the influence of only the other resonance state allowing an 
analysis in a very transparent manner. 
This  property qualifies the energy dependent Fano parameter
(\ref{li6}) for the description of resonances in the overlapping
regime. We underline, however, that the
$\tilde q_k(E)$ are not suitable for 
the parametrization of the cross section around a double pole of the $S$
matrix due to the quadratic term appearing in (\ref{brwi6}).
Here, the cross section can be parametrized as 
$\sigma(E)=16\cos^2\delta[(\tilde
q_d(E)+\varepsilon_d)^2/(\varepsilon_d^2+1)^2]$ with $\tilde
q_d(E)=\case{1}{2}(1-\varepsilon_d^2)\tan\delta$ and $\varepsilon_1 \to
\varepsilon_2 \equiv \varepsilon_d$. 

It would be interesting to
look for the broad resonance state(s) and to study its (their) influence on
other narrow resonances in detail. According to our study, the
broad resonance may exist in the arm or it may coexist with the narrow
resonances  in the quantum dot itself.  The last possibility
can not be excluded theoretically. Quite on the contrary, it
corresponds to the results obtained for other small quantum systems
where resonance states (eigenstates of the effective Hamiltonian)
of very different lifetimes are known to coexist
\cite{rep}.  An experimental study of this question would allow
to see  generic features of open quantum systems also in quantum dots.
Furthermore,  results from the interference between two
resonance states in the very neighbourhood of a double pole of the $S$ matrix
(that can be studied  in a symmetrical device with two quantum dots)
will prove the resonance scenario described by an effective Hamiltonian as
discussed in the present paper.

\vspace{.2cm}

Summarizing, we conclude that the line shape of narrow resonances
in the overlapping regime contains information on generic
properties of open quantum systems. One of these properties is the
existence of different time scales that are involved in the
eigenvalues and eigenfunctions of the effective Hamiltonian of the
system.  A control by external parameters  allows to trace their
formation by opening the system and to study their interesting
interplay. A direct experimental study of the generic resonance
features in, e.g., atomic nuclei is difficult due to the strong
residual interaction between the participating particles in
nuclei.  Quantum dots are  much more suitable for such a study due
to the flexibility in controlling the system by means of different
external parameters.  Further experimental as well as theoretical
studies are  highly desirable for both a better understanding of
generic properties of open quantum systems and the construction of
quantum dots with special properties.

\vspace{.2cm}

We are indebted to J.M. Rost for valuable discussions.
A.I.M. gratefully acknowledges the hospitality of the Max-Planck-Institut
f\"ur Physik komplexer Systeme.

\end{document}